\documentclass[preprint, 5p,times]{elsarticle}

\usepackage[utf8]{inputenc}

\usepackage{amssymb}
\usepackage{amsmath,amssymb,amsfonts}
\usepackage{graphicx}
\usepackage{textcomp}
\usepackage{xcolor}
\usepackage{xurl}
\usepackage{tabularx}
\usepackage{multirow}
\usepackage{xcolor}%
\usepackage{textcomp}%
\usepackage{manyfoot}%
\usepackage{booktabs}%
\usepackage{listings}%
\usepackage{arydshln}
\usepackage{adjustbox}
\usepackage{algorithm}
\usepackage{algpseudocode}
\usepackage{comment}
\usepackage{tabularray}
\usepackage{balance}
\usepackage{threeparttable}
\usepackage{url}

\makeatletter
\newlength{\fslength}
\newcommand{\funnyP}{%
    \setlength{\fslength}{\f@size pt}%
    \mathord{\reflectbox{$\mathrm{P}$}}\mkern-6mu\mathord{\mathrm{P}}%
}
\newcommand{\funnyPbar}{%
    \mathord{\reflectbox{$\bar{\mathrm{P}}$}}\mkern-6mu\mathord{\bar{\mathrm{P}}}%
}

\journal{Future Generation Computer Systems}

\myfooter[L]{\parbox{\textwidth}{%
  \raggedright
  © $<$2025$>$. This manuscript version is made available under the CC-BY-NC-ND 4.0 license\\
  \texttt{https://creativecommons.org/licenses/by-nc-nd/4.0/}\\
  The published journal article is available at: \texttt{https://doi.org/10.1016/j.future.2025.107838}
}}

\begin{document}

\begin{frontmatter}



\title{Analyzing the Performance Portability of SYCL across CPUs, GPUs, and Hybrid Systems with SW Sequence Alignment} 





\author{Manuel Costanzo\fnref{fn1}}
\ead{mcostanzo@lidi.info.unlp.edu.ar}
\author{Enzo Rucci\fnref{fn1,fn2}}
\ead{erucci@lidi.info.unlp.edu.ar}
\author{Carlos García-Sánchez\fnref{fn3}}
\ead{garsanca@ucm.es}
\author{Marcelo Naiouf\fnref{fn1}}
\ead{mnaiouf@lidi.info.unlp.edu.ar}
\author{Manuel Prieto-Matías\fnref{fn3}}
\ead{mpmatias@ucm.es}

\affiliation[fn1]{organization={III-LIDI, Facultad de Informática},
            addressline={50 y 120}, 
            city={La Plata},
            postcode={1900}, 
            state={Buenos Aires},
            country={Argentina}}

\affiliation[fn2]{organization={Comisión de Investigaciones Científicas (CIC)},
            city={La Plata},
            state={Buenos Aires},
            country={Argentina}}

\affiliation[fn3]{organization={Dpto. Arquitectura de Computadores y Automática, Universidad Complutense de Madrid},
            city={Madrid},
            postcode={28040},
            country={España}}

\begin{abstract}
The high-performance computing (HPC) landscape is undergoing rapid transformation, with an increasing emphasis on energy-efficient and heterogeneous computing environments. This comprehensive study extends our previous research on SYCL's performance portability by evaluating its effectiveness across a broader spectrum of computing architectures, including CPUs, GPUs, and hybrid CPU-GPU configurations from NVIDIA, Intel, and AMD. Our analysis covers single-GPU, multi-GPU, single-CPU, and CPU-GPU hybrid setups, using  two common, bioinformatic applications as a case study. The results demonstrate SYCL's versatility across different architectures, maintaining comparable performance to CUDA on NVIDIA GPUs while achieving similar architectural efficiency rates on AMD and Intel GPUs in the majority of cases tested. SYCL also demonstrated remarkable versatility and effectiveness across CPUs from various manufacturers, including the latest hybrid architectures from Intel.  Although SYCL showed excellent functional portability in hybrid CPU-GPU configurations, performance varied significantly based on specific hardware combinations. Some performance limitations were identified in multi-GPU and CPU-GPU configurations, primarily attributed to workload distribution strategies rather than SYCL-specific constraints. These findings position SYCL as a promising unified programming model for heterogeneous computing environments, particularly for bioinformatic applications.
\end{abstract}



\end{frontmatter}

\section{Introduction}
\label{sec:intro}

The parallel processing landscape is undergoing rapid transformation, with an increasing emphasis on energy-efficient and heterogeneous computing environments~\cite{navaux23}. GPUs are now the dominant accelerator in this field, offering incomparable performance for parallel workloads. Still, the roles of CPUs and the potential of hybrid CPU-GPU systems are also actively explored in computational research. On the one hand, GPU-based Systems on Chip (SoC) are becoming more popular every day. On the other hand, hybridization now extends to CPU architectures, with the emergence of heterogeneous designs like Intel's Raptor Lake and Meteor Lake. Depending on the workload, these processors incorporate up to three different types of cores on the same die to optimize performance and energy efficiency. This architectural heterogeneity at the system and processor levels reflects the larger market dynamics and the ongoing evolution of computing architectures.


As of 2024, NVIDIA maintains its dominance in the GPU market, particularly in the discrete GPU (dGPU) segment, where it possesses an 88\% market share. AMD has demonstrated significant progress, capturing 12\% of the market, while Intel experienced a decline, with its market share dropping to less than 1\%~\cite{shilov24}. In the integrated graphics (iGPU) sector, Intel retains a substantial advantage with a 68\% market share, followed by NVIDIA (18\%) and AMD (14\%). This market structure highlights the importance of considering diverse architectures in GPU-based computing research~\cite{norem24}.

Concerning parallel programming, the proliferation of diverse computing architectures (including CPUs, GPUs, and hybrid CPU-GPU systems), has created a pressing need for unified programming models that can effectively harness their collective potential. In this context, SYCL has emerged as a promising solution that offers a cross-platform abstraction layer that facilitates code portability across various hardware architectures. By enabling developers to write high-performance code that can be seamlessly executed on a wide range of devices, SYCL plays a crucial role in addressing the challenge of performance portability in heterogeneous computing environments~\cite{SYCL}.

This article evolves from our systematic exploration of SYCL's capabilities in bioinformatics applications across heterogeneous computing platforms. In our initial work~\cite{Costanzo2022IWBBIO}, we focused on migrating a CUDA-based, protein database search software (SW\#db) to SYCL using Intel's oneAPI ecosystem. This study demonstrated the tool's effectiveness for code migration and it defined the basic cross-GPU-architecture portability of the migrated code, achieving comparable or slightly better efficiency rates compared to the native CUDA implementation. Building upon these findings, in~\cite{costanzo23} we expanded the study to evaluate SYCL's performance portability across different GPU architectures, including single and multi-GPU configurations from different vendors. This research revealed that while CUDA and SYCL versions achieved similar performance on NVIDIA devices, SYCL reached higher architectural efficiency rates in most test cases from the other GPU vendors. Finally, in~\cite{supercompcost23} we completed the SYCL-migration process of SW\# biological suite  to support pairwise alignment and different alignment algorithms. Moreover, its performance in a wide variety of biological test scenarios was analyzed and cross-SYCL-implementation portability was verified on several GPUs and some CPUs from a functional perspective.

This work extends our performance portability study to include pairwise sequence alignment and (pure and hybrid) CPUs and CPU-GPU systems. The key contributions of this paper are:

\begin{itemize}
    \item An extension of our previously adapted performance model from~\cite{lan2017swhybrid}, to encompass pure and hybrid CPU architectures and also combined CPU-GPU systems.
    
    \item An extension of our previous work on the functional and performance portability of \textit{SW\#} applications across different GPU architectures to now include pairwise sequence alignment and Intel and AMD CPUs, as well as hybrid CPU-GPU configurations. 
    On the one side, by investigating the performance characteristics of SYCL on these diverse platforms, this study aims to provide valuable insights into its potential as a unified programming model for the increasingly complex landscape of modern computing. On the other side, by expanding the study to include CPUs and hybrid systems, this research seeks to offer a more comprehensive understanding of how SYCL-based applications perform across the full spectrum of available computational resources.
    
    \item An analysis of the trade-offs and synergies between CPU and GPU computations in hybrid configurations for two common bioinformatic workloads.
\end{itemize}

The remainder of this paper is structured as follows: Section~\ref{sec:back} presents the background; Section~\ref{sec:imps} details the case study applications and the expanded performance model; Section~\ref{sec:results} presents our findings on functional and performance portability; Section~\ref{sec:relworks} discusses related work; and Section~\ref{sec:conc} presents the conclusions and future research directions.

\section{Background}
\label{sec:back}

\subsection{CPUs, GPUs, and Programming Models for Heterogeneous Computing}
The evolution of high-performance computing (HPC) has been significantly influenced by the emergence of general-purpose GPU programming. In 2007, NVIDIA introduced CUDA~\cite{CUDA_handson} alongside the Tesla GPU architecture, enabling non-graphics applications to harness GPU computational power. Although CUDA quickly became a prominent low-level programming model for GPU computing, its proprietary nature and exclusive compatibility with NVIDIA hardware led to the development of more versatile alternatives, such as OpenCL~\cite{OpenCL}. OpenCL offers a level of abstraction comparable to that of CUDA while supporting multiple devices and vendors.

The field has also expanded to include multicore CPUs and hybrid GPU-CPU systems, accompanied by the development of high-level programming initiatives such as OpenMP~\cite{OpenMP}, OpenACC~\cite{OpenACC, OpenACC_book}, and SYCL~\cite{SYCL}. Although CUDA is limited to NVIDIA hardware, and OpenCL provides broad coverage but is challenging to learn and costly to implement, SYCL has emerged as a promising solution for heterogeneous computing.

SYCL~\cite{SYCL} enables developers to write code for various processors, including CPUs, GPUs, and hybrid systems, using standard ISO C++. It incorporates host and device code in a single source file and supports multiple acceleration APIs, such as OpenCL, facilitating seamless integration with lower-level code across different hardware architectures.

Several SYCL implementations are now available, with Intel's oneAPI~\cite{DPCPP} standing out as a mature developer suite. OneAPI eliminates the need for separate code bases for host and device, which is required in OpenCL, and supports multiple programming languages and tools for different architectures. Data Parallel C++ (DPC++), oneAPI's core language~\cite{DPCPP}, integrates the SYCL and OpenCL standards without additional extensions, facilitating the development of applications that can efficiently utilize the full spectrum of available computational resources.

\subsection{Smith-Waterman Sequence Alignment}
\label{sec:SW-Algorithm}

A key process in bioinformatics and computational biology is sequence alignment, which aims to identify regions of similarity between sequences to uncover their structural, functional, and evolutionary relationships~\cite{PSAreview}. The Smith-Waterman (SW) algorithm is commonly employed to determine the optimal local alignment between two sequences~\cite{Smith1981}. This technique relies on dynamic programming and is highly sensitive, as it examines all possible alignments between the sequences.


Given two sequences $Q$ and $D$ of length $|Q|=m$ and $|D|=n$, the recurrence relations for the SW algorithm with the modification of Gotoh~\cite{gotoh81} are defined as follows:


\begin{equation}
	H_{i,j}=max
	\begin{cases}
		0\\
		H_{i-1,j-1}+SM(Q[i],D[j])\\
		E_{i,j}\\
		F_{i,j}
	\end{cases}
	\label{eq:sw2}
\end{equation}

\begin{equation}
	E_{i,j}=max
	\begin{cases}
		H_{i,j-1} - G_{o}\\
		E_{i,j-1} - G_{e}
	\end{cases}
	\label{eq:sw3}
\end{equation}
	
\begin{equation}
	F_{i,j}=max
	\begin{cases}
		H_{i-1,j} - G_{o}\\
		F_{i-1,j} - G_{e}
	\end{cases}
	\label{eq:sw4}
\end{equation}

The similarity score $H_{i,j}$ is calculated to identify a common subsequence; $H_{i,j}$ contains the score to align the prefixes $Q[1..i]$ and $D[1..j]$. Moreover, $E_{i,j}$ and $F_{i,j}$ correspond to the scores of prefix $Q[1..i]$ and $D[1..j]$ aligned to a gap, respectively. \emph{SM} denotes the \emph{scoring matrix} and defines the match/mismatch scores between the residues. Lastly, $G_{o}$ and $G_{e}$ refer to the gap open and gap extension penalties, respectively. 

Firstly, $H$, $E$ and $F$ must be initialized with 0 when $i = 0$ or $j = 0$. Then, the recurrences should be calculated with $1 \leq i \leq m$ and $1 \leq j \leq n$. The highest value in the matrix $H$ ($S$) corresponds to the optimal local alignment score between $Q[1..i]$ and $D[1..j]$. If required, the optimal local alignment is finally obtained by following a traceback procedure whose starting point is $S$. From a computational perspective, it is important to highlight the dependencies of any $H$ element. Any cell can be calculated only after the values of the upper, left, and upper-left neighbors are known; imposing restrictions on how \emph{H} can be processed.

\begin{figure}[!t]
 \centering
      \includegraphics[width=0.75\columnwidth]{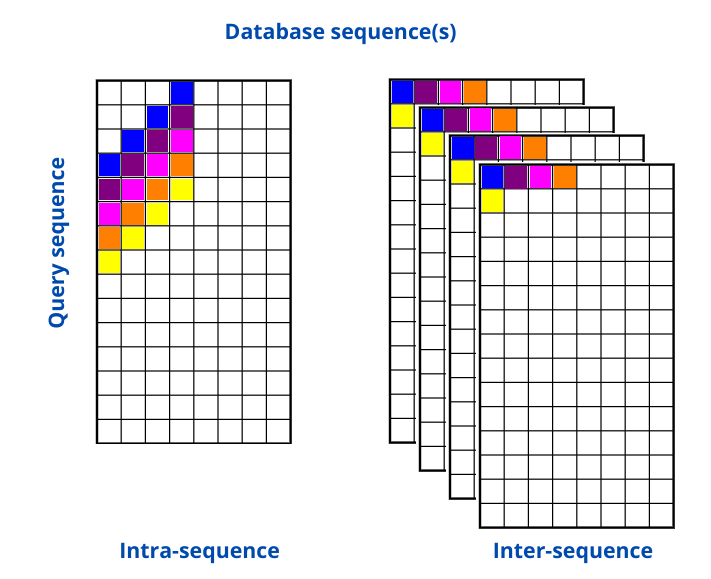}
     \caption{ Parallelization approaches in similarity matrix computations (adapted from~\cite{swipe11}). Each color indicates the cells that can be computed together in a SIMD manner.}
     \label{fig:schemes}
 \end{figure}

 SW is usually employed to compute: (a) pairwise alignments of two very long DNA sequences, and (b) database similarity search (one-to-many), involving shorter protein sequences. As SW runs in quadratic time to compute optimal alignment, widely adopted parallelization approaches are \textit{intra-sequence (a single matrix is calculated and all Processing Elements (PEs) work collaboratively) and \textit{inter-sequence} (simultaneously calculating multiple matrices without communication between the PEs). Fig.~\ref{fig:schemes} illustrates both approaches.}

\subsection{Performance portability}

According to Penycook et al.~\cite{performance_portability_paper}, \textit{performance portability} refers to \textit{"A measurement of an application’s performance efficiency for a given problem that can be executed correctly on all platforms in a given set"}. These authors define two different performance efficiency metrics:  \textit{architectural efficiency} and \textit{application
efficiency}. The former denotes the capacity of an application to effectively utilize hardware resources, measured as a proportion of the theoretical peak performance. The latter signifies the application's ability to select the most suitable implementation for each platform, representing a fraction of the highest observed performance achieved.

The performance portability metric presented by Penycook et al.~\cite{performance_portability_paper} was later reformulated by Marowka~\cite{performance_portability_paper_rev} to address some of its flaws. Formally, for a given set of platforms \textit{H} from the same architecture class, the portability $\funnyPbar$ of a case study application $\alpha$ to solve the problem $p$ is:

$$
\funnyPbar(\alpha, p, H)= \begin{cases}\frac{\sum_{i \in H}{e_i(\alpha, p)}}{|H|} & \text { if } i \text { is supported } \forall i \in H \\ \text{not applicable (NA)} & \text { otherwise }\end{cases}
$$

where $e_i(\alpha, p)$ corresponds to the performance efficiency of case-study application $\alpha$ solving problem \textit{p} on the platform \textit{i}.

The \textit{performance portability} concept emphasizes the capability to write code that can efficiently utilize the available computing resources, such as CPUs, GPUs, or specialized accelerators, while maintaining high performance regardless of the specific hardware configuration. With performance portability, developers can write code once and have it deliver optimal performance on various target platforms. This eliminates the need for extensive manual code optimizations or platform-specific modifications, reducing development time and effort.

\label{sec:perf-port}

\section{Case-Study Applications and Performance Model}
\label{sec:imps}

\subsection{Case-Study Applications}
\label{sec:case-study}
Two GPU-accelerated implementations of SW sequence alignment were considered for the performance portability evaluation:
\begin{itemize}
    \item \texttt{CUDA}: The \textit{SW\#} framework represents a CUDA-implemented solution for biological sequence alignment that prioritizes memory efficiency. Operating as both a standalone program and an integrated library, it enables protein and DNA sequence comparisons through pairwise alignment and database searches. The framework provides customizable alignment parameters, including Smith-Waterman algorithm options, gap penalty specifications, and scoring matrix selection. Through  CPU-GPU coordination, \textit{SW\#} achieves good performance by implementing a dynamic load balancing system that assigns tasks based on sequence length characteristics. The parallelization architecture primarily leverages GPU capabilities through both inter-task and intra-task approaches. GPU processing follows a dual-kernel strategy: shorter database sequences are processed via inter-task parallelism in a \textit{short kernel}, while longer sequences utilize intra-task parallelism in a \textit{long kernel}. For multi-GPU configurations, \textit{SW\#} employs an adaptive execution protocol. When the GPU count exceeds query sequence numbers, all devices synchronously process different database segments for a single query. Alternatively, with more queries than GPUs, each GPU independently processes distinct queries against the complete database~\cite{swsharp,swsharpdb}.

    \item \texttt{SYCL}: this code is based on the implementation presented in the paper~\cite{supercompcost23}, representing a SYCL equivalent. The migration of the \textit{SW\#} suite was performed using \texttt{dpct} (the Data Parallel Compatibility Tool available in the oneAPI suite) and some hand-coding modifications. The resulting implementation is pure SYCL code without device-specific optimizations or native code paths, ensuring that all computations—whether on GPUs or CPUs—are executed through SYCL. This approach maintains complete portability across different architectures while allowing for a fair comparison of SYCL's performance in various computing environments.

\end{itemize}

\subsection{Performance Model}
\label{sec:perf-model}

Peak theoretical hardware performance must be estimated for all selected devices in this study to compute the performance portability metric. This step requires considering both hardware and algorithm features. In our previous work~\cite{costanzo23}, we have adapted the performance model from~\cite{lan2017swhybrid}
to the \textit{SW\#} algorithm and several GPUs.  For the present investigation, we extend this adapted model to encompass CPUs from various vendors. In fact, the presented equations can serve as a generic framework for estimating the theoretical peak performance of multiple devices, including CPUs, GPUs, and even FPGAs. The computing capability of different devices can be estimated using Eq.~\ref{eq:cc-gpu}:
\begin{equation}
\label{eq:cc-gpu}
    Capability = Clock\_Rate \times Throughput \times Lanes
\end{equation}
where \textit{Clock\_rate} denotes the device's clock frequency,  \textit{Throughput} represents the number of instructions the device can execute per clock cycle, and \textit{Lanes} indicates the quantity of SIMD vector lanes. Subsequently, it is imperative to enumerate the instructions issued during each cell update of the similarity matrix. In the sequence alignment context, the predominant performance metric is Cell Updates Per Second (CUPS), representing the time for a complete computation of one cell in the similarity matrix (\textit{H). The GCUPS (billion CUPS) value is calculated by:
\begin{equation}
\label{eq:gcups}
    GCUPS = \frac{Q \times D}{t \times 10^9}
\end{equation}
where $|Q|$ is the total number of residues in the query sequence, $|D|$ is the total number of residues in the database and \textit{t} is the runtime in seconds}~\cite{RucciBDAG2016}. Consequently, the theoretical peak GCUPS of any device can be modeled using Equation~\ref{eq:theo-peak}:
\begin{equation}
\label{eq:theo-peak}
    Theo\_peak = \frac{Capability}{Instruction\_count\_one\_cell\_update}
\end{equation}
 Table~\ref{tab:GCUPS-peak} provides a comprehensive summary of the theoretical peak performance for selected GPUs, calculated using Equation~\ref{eq:theo-peak}.  Additionally, the performance metrics for CPUs are elucidated in Table~\ref{tab:GCUPS-peak-cpu}. Further detailed analysis and discussion are presented in the subsequent sections of this paper.

\subsubsection{SW\# core instructions}

\textit{SW\#} computes the similarity matrix using 32-bit integers and performs 12 instructions per cell update. Algorithm~\ref{alg:core} presents the snippet of cell update in similarity matrix as in Eq.~\ref{eq:sw2}, Eq.~\ref{eq:sw3}, and Eq.~\ref{eq:sw4}. Just adding, subtracting, and maximum instructions are required to perform a single-cell update.

\begin{algorithm}

\begin{algorithmic}[1]
\State $E^{1} = E_{l} - G_{e}$
\Comment{$E_{l}$: E of its left neighbor}
\State $E^{2} = H_{l} - G_{o}$
\Comment{$H_{l}$: H of its left neighbor}
\State $E = max(E^{1},E^{2})$
\State $F^{1} = F_{u} - G_{e}$
\Comment{$F_{u}$: F of its upper neighbor}
\State $F^{2} = H_{u} - G_{o}$
\Comment{$H_{u}$: H of its upper neighbor}
\State $F = max(F^{1},F^{2})$
\State $H = H_{ul} + SM$ \Comment{$H_{ul}$: H of its upper-left neighbor}
\State $H = max(H,E)$
\State $H = max(H,F)$
\State $H = max(H,0)$
\State $A = H$ \Comment $A$: an auxiliary variable
\State $S = max(H,S)$ \Comment $S$: optimal score
\end{algorithmic}
\caption{ Core instructions per cell update in similarity matrix}
\label{alg:core}
\end{algorithm}

\subsubsection{Architectural features on NVIDIA GPUs}

The \# Cores in an NVIDIA GPU refers to the number of Streaming Multiprocessors. CUDA does not strictly follow a SIMD execution model but it adopts a similar one denoted as the SIMT model. A \textit{warp} comprises a group of 32 threads that execute the same instruction stream. According to~\cite{lan2017swhybrid}, ``a \textit{warp} in SIMT is equivalent to a \textit{vector} in SIMD, and a \textit{thread} in SIMT is equivalent to a \textit{vector lane} in SIMD''. The instruction throughput depends on the CUDA Compute Capability (CC) of each NVIDIA GPU~\footnote{\url{https://docs.nvidia.com/cuda/cuda-c-programming-guide/#maximize-instruction-throughput}}.

\subsubsection{Architectural features on AMD GPUs}

In the RDNA2.0 architecture, the \# Cores parameter corresponds to the number of Compute Units (CUs), which are organized in pairs to form Workgroup Processors (WPs). AMD employs the terms \textit{wavefront} and \textit{work-item} as analogs to NVIDIA's \textit{warp} and \textit{thread}, respectively. While RDNA2.0 supports wavefront sizes of both 32 and 64 work items, the former has priority over the latter. Each CU encompasses two SIMD32 vector units, enabling the execution of 64 add/subtract/max instructions per cycle (\texttt{Int32}). Consequently, the instruction throughput per work item is quantified as 2.

\subsubsection{Architectural features on Intel GPUs}

In the dGPU segment, Intel's GPU design philosophy diverges significantly from that of NVIDIA and AMD. The fundamental building block of the Intel Xe microarchitecture is the Xe Core, each comprising 16 Xe Vector Engines (XVEs)~\footnote{Also referred to as Execution Units (EUs)}, capable of executing 8 add/subtract/max instructions per cycle (\texttt{Int32}). In the context of the proposed model, Xe Cores and XVEs correspond to \# Cores and \# Lanes, respectively.

In the iGPU segment, both Gen9 and Gen12 microarchitectures exhibit similar design principles, primarily differentiated by the number of computational resources. These microarchitectures utilize the Subslice as their fundamental block, each containing 8 Execution Units (EUs) that can process 8 add/subtract/max instructions per cycle (\texttt{Int32}). Within the framework of the proposed model, Subslices and EUs are analogous to \# Cores and \# Lanes, respectively.

\subsubsection{Architectural features on CPUs}

For Intel and AMD CPUs, \# Cores represents the number of processor cores, each capable of independently executing instructions, which is crucial for performance in parallel applications. The \# Lanes is defined by the number of Vector Processing Units (VPUs) and their supported vector width/instruction set (such as SSE, AVX, AVX-512), which allows a core to perform multiple operations simultaneously. The instruction throughput depends on the operation that is being performed.

Since intensive use of VPUs leads to increased heat dissipation and power consumption, modern processors incorporate various technologies that reduce the clock frequency to counteract their impact~\cite{Rutter}. For this reason, the operating frequency when all cores perform intensive computation is usually the base frequency or a slightly higher one, particularly on the server segment. For this work, we have monitored the \textit{SW\#} execution on each CPU device to verify the information provided.

Finally, a caveat must be made for hybrid CPU architectures (such as Intel's Alder Lake)~\cite{AlderLakes}. As these architectures present up to three types of cores (P-cores, E-cores, and LP E-cores), the maximum theoretical performance is obtained by summing up the partial maximum performances.

\begin{table*}[t!]
\resizebox{\textwidth}{!}{%
\begin{tabular}{|l|cccccc|cccc|cc|}
\hline
\textbf{Vendor} &
  \multicolumn{6}{c|}{NVIDIA} &
  \multicolumn{4}{c|}{Intel} &
  \multicolumn{2}{c|}{AMD} \\ \hline
\textbf{Model} &
  \multicolumn{1}{c|}{GTX 980} &
  \multicolumn{1}{c|}{GTX 1080} &
  \multicolumn{1}{c|}{RTX 2070} &
  \multicolumn{1}{c|}{V100} &
  \multicolumn{1}{c|}{RTX 3070} &
  RTX 3090 &
  \multicolumn{1}{c|}{Arc A770} &
  \multicolumn{1}{c|}{UHD 630} &
  \multicolumn{1}{c|}{UHD 770} &
  Xe-LPG 128EU &
  \multicolumn{1}{c|}{RX 6700 XT} &
  \multicolumn{1}{c|}{Vega 6}
  \\ \hline
\textbf{Type} &
  \multicolumn{6}{c|}{Discrete} &
  \multicolumn{1}{c|}{Discrete} &
  \multicolumn{3}{c|}{Integrated} &
  \multicolumn{1}{c|}{Discrete} &
  \multicolumn{1}{c|}{Integrated}  \\ \hline
\textbf{Architecture} &
  \multicolumn{1}{c|}{\begin{tabular}[c]{@{}c@{}}Maxwell \\ (CC 5.2)\end{tabular}} &
  \multicolumn{1}{c|}{\begin{tabular}[c]{@{}c@{}}Pascal \\ (CC 6.1)\end{tabular}} &
  \multicolumn{1}{c|}{\begin{tabular}[c]{@{}c@{}}Turing \\ (CC 7.5)\end{tabular}} &
  \multicolumn{1}{c|}{\begin{tabular}[c]{@{}c@{}}Volta \\ (CC 7.0)\end{tabular}} &
  \multicolumn{1}{c|}{\begin{tabular}[c]{@{}c@{}}Ampere \\ (CC 8.6)\end{tabular}} &
  \begin{tabular}[c]{@{}c@{}}Ampere \\ (CC 8.6)\end{tabular} &
  \multicolumn{1}{c|}{Xe} &
  \multicolumn{1}{c|}{Gen 9.5} &
  \multicolumn{1}{c|}{Gen 12} &
  Xe &
  \multicolumn{1}{c|}{RDNA 2.0} &
  \multicolumn{1}{c|}{GCN5} 
  \\ \hline
\textbf{\# Cores} &
  \multicolumn{1}{c|}{16} &
  \multicolumn{1}{c|}{20} &
  \multicolumn{1}{c|}{36} &
  \multicolumn{1}{c|}{80} &
  \multicolumn{1}{c|}{46} &
  82 &
  \multicolumn{1}{c|}{32} &
  \multicolumn{1}{c|}{3} &
  \multicolumn{1}{c|}{4} &
  8 &
  \multicolumn{1}{c|}{40} &
  \multicolumn{1}{c|}{6} 
  \\ \hline
\textbf{\# Lanes} &
  \multicolumn{7}{c|}{32} &
  \multicolumn{1}{c|}{16} &
  \multicolumn{1}{c|}{8} &
  \multicolumn{1}{c|}{16} &
  \multicolumn{2}{c|}{32} 
  \\ \hline
\textbf{\begin{tabular}[c]{@{}c@{}}Inst. throu.\end{tabular}} &
  \multicolumn{1}{c|}{4/2} &
  \multicolumn{1}{c|}{4/2} &
  \multicolumn{4}{c|}{2} &
  \multicolumn{4}{c|}{8} &
  \multicolumn{2}{c|}{2} 
  \\ \hline
\textbf{Clock (MHz)} &
  \multicolumn{1}{c|}{1216} &
  \multicolumn{1}{c|}{1733} &
  \multicolumn{1}{c|}{1620} &
  \multicolumn{1}{c|}{1380} &
  \multicolumn{1}{c|}{1725} &
  1695 &
  \multicolumn{1}{c|}{2400} &
  \multicolumn{1}{c|}{1200} &
  \multicolumn{1}{c|}{1650} &
  2250 &
  \multicolumn{1}{c|}{2581}&
  \multicolumn{1}{c|}{1100}
  \\ \hline
\textbf{\begin{tabular}[c]{@{}l@{}}Peak (GCUPS)\end{tabular}} &
  \multicolumn{1}{c|}{155.6} &
  \multicolumn{1}{c|}{277.3} &
  \multicolumn{1}{c|}{311} &
  \multicolumn{1}{c|}{588.8} &
  \multicolumn{1}{c|}{423.2} &
  741.2 &
  \multicolumn{1}{c|}{819.2} &
  \multicolumn{1}{c|}{19.2} &
  \multicolumn{1}{c|}{35.2} &
  192 &
  \multicolumn{1}{c|}{550.6} &
  \multicolumn{1}{c|}{35.2}
  \\ \hline
\multicolumn{11}{l}{\begin{tabular}[c]{@{}l@{}}The instruction throughput for GTX 980 and GTX 1080 is 4 for add/subtract and 2 for max/min. The core instructions include 5 \\ add/subtract and 6 max. Thus, the equivalent throughput is 3.\end{tabular}} \\
\multicolumn{11}{l}{The core instruction count for each cell update is 12.}  
\end{tabular}
}

\caption{GPU specifications and their theoretical peak performance in terms of GCUPS}
\label{tab:GCUPS-peak}
\end{table*}

\begin{table*}[t!]
\resizebox{\textwidth}{!}{%
\centering
\begin{threeparttable}
\small
\begin{tabular}{|l|c|c|c|c|c|c|c|c|c|}
\hline
\textbf{Vendor} & \multicolumn{8}{c|}{Intel} & AMD \\
\hline
\textbf{Model} & \begin{tabular}[c]{@{}c@{}}Core\\i5-7400\end{tabular} & 
\begin{tabular}[c]{@{}c@{}}Core\\i5-10400F\end{tabular} & 
\begin{tabular}[c]{@{}c@{}}Xeon\\E5-1620 v3\end{tabular} & 
\begin{tabular}[c]{@{}c@{}}Xeon\\E5-2695 v3\end{tabular} & 
\begin{tabular}[c]{@{}c@{}}Xeon\\Gold 6138\end{tabular} & 
\begin{tabular}[c]{@{}c@{}}Core\\i9-9900K\end{tabular} & 
\begin{tabular}[c]{@{}c@{}}Core\\i9-13900K\end{tabular} & 
\begin{tabular}[c]{@{}c@{}}Core\\ Ultra 9-185H	\end{tabular} & 
\begin{tabular}[c]{@{}c@{}}Ryzen 3\\ 5300U\end{tabular} \\
\hline
\textbf{Segment} & \multicolumn{2}{c|}{Desktop} & \multicolumn{3}{c|}{Server} & \multicolumn{2}{c|}{Desktop} & \multicolumn{2}{c|}{Mobile} \\
\hline
\textbf{Architecture} & Kaby Lake & Comet Lake & \begin{tabular}[c]{@{}c@{}}Sandy\\Bridge-E\end{tabular} & Haswell & Skylake & \begin{tabular}[c]{@{}c@{}}Coffee\\Lake-R\end{tabular} & \begin{tabular}[c]{@{}c@{}}Raptor\\Lake-S\end{tabular} & \begin{tabular}[c]{@{}c@{}}Meteor\\Lake-S\end{tabular}& Lucienne \\
\hline
\textbf{\# Cores} & 4 & 6 & 4 & 14 & 40 & 8 & 8/16 & 6/8/2 & 4 \\
\hline
\textbf{SIMD set} & AVX2 & AVX2 & AVX2 & AVX2 & AVX-512 & AVX2 & AVX2 & AVX2 & AVX2 \\
\hline
\textbf{\# Lanes} & \multicolumn{4}{c|}{8} & 16 & \multicolumn{4}{c|}{8} \\
\hline
\textbf{Inst. throu.} & \multicolumn{9}{c|}{1} \\
\hline
\textbf{Clock (MHz)} & 3300 & 4000 & 3500 & 1900 & 1900 & 4700 & 5500/4300 & 5100/3800/2500 & 3600 \\
\hline
\textbf{Peak (GCUPS)} & 8.8 & 16 & 9.3 & 35.5 & 101.3 & 25.1 & 75.2 & 44 & 9.6 \\
\hline
\end{tabular}
\end{threeparttable}
}

\begin{tablenotes}\small
  \item[a] The Intel Core i9-13900K CPU features 8 P-cores and 16 E-cores.
  \item[b] The Intel Core Ultra 9-185H CPU features 6 P-cores, 8 E-cores, and 2 LP E-cores.
\end{tablenotes}

\caption{CPU specifications and their theoretical peak performance in terms of GCUPS}
\label{tab:GCUPS-peak-cpu}
\end{table*}

\section{Experimental Results}
\label{sec:results}

\subsection{Experimental Design}
The experiments were carried out on 12 GPUs, including 6 NVIDIA dGPUs, 1 AMD dGPU, 1 AMD iGPU, 3 Intel iGPUs, and 1 Intel dGPU. In addition, 8 Intel CPUs (from different segments) and 1 AMD CPU were included in the study. The specific details of these platforms can be found in Table~\ref{tab:GCUPS-peak}. The oneAPI and CUDA versions used were \texttt{2022.1.0} and \texttt{11.7}, respectively. For both CUDA and SYCL, the optimization flag \texttt{-O3} was used during compilation. To run SYCL code on NVIDIA and AMD GPUs, several modifications had to be made to the build process, as SYCL is not supported by default on these platforms\footnote{\url{https://intel.github.io/llvm-docs/GetStartedGuide.html}} but Codeplay recently announced free binary plugins\footnote{\url{https://codeplay.com/portal/blogs/2022/12/16/bringing-nvidia-and-amd-support-to-oneapi.html}} to support it. After these modifications, it was possible to run DPC++ code on both NVIDIA and AMD GPUs using the Clang++ compiler (\texttt{16.0}).

To ensure that the CPU executes only SYCL code and avoid the use of SW\# ``native'' CPU code, the flag \texttt{T=0} was used in all tests. This setting forces all sequence alignments to be thoroughly computed using SYCL, allowing for a fair evaluation of SYCL performance on the CPU.

Performance evaluation was carried out using real biological data for the two application benchmarks:

\begin{itemize}
    \item Protein database search: searching 20 query protein sequences against the well-known Environmental Non-Redundant database (Env. NR) (\texttt{2021\_04 Release}), which contains 995210546 amino acid residues in 4789355 sequences, with a maximum length of 16925. Query sequences were selected from the Swiss-Prot database~\footnote{Swiss-Prot: ~\url{https://www.uniprot.org/downloads}}, with lengths ranging from 144 to 5478. The access numbers for these queries are: P02232, P05013, P14942, P07327, P01008, P03435, P42357, P21177, Q38941, P27895, P07756, P04775, P19096, P28167, P0C6B8, P20930, P08519, Q7TMA5, P33450, and Q9UKN1. Besides, \textit{SW\#} was configured with \texttt{BLOSUM62} as the substitution matrix, and 10/2 as the penalty for insertion/extension. 
    \item Pairwise alignment: computing five different DNA alignments. Table~\ref{tab:dna} presents the accession numbers and sizes of the DNA sequences used. The score parameters were configured as +1 for match, -3 for mismatch, -5 for gap open, and -2 for gap extension.
\end{itemize}

\begin{table}[htbp]
\begin{tabular*}{\columnwidth}{@{\extracolsep\fill}ccccl}
\toprule%
\multicolumn{2}{@{}c@{}}{Query sequence} & \multicolumn{2}{@{}c@{}}{Database sequence} \\\cmidrule{1-2}\cmidrule{3-4}%
Accession & Size & Accession & Size & \begin{tabular}[c]{@{}c@{}}Matrix Size\\ (cells)\end{tabular} \\
\midrule
NC\_000898 & 162K &  NC\_007605 &  172K & 28M  \\
NC\_003064.2 & 543K &  NC\_000914.1 &  536K & 291M  \\
CP000051.1 & 1M &  AE002160.2 &  1M & 1G  \\
BA000035.2 & 3M &  BX927147.1 &  3M & 9G  \\
NC\_005027.1 & 7M &  NC\_003997.3 &  5M & 35G  \\
\bottomrule
\end{tabular*}
\caption{DNA sequence information used in the tests}\label{tab:dna}
\end{table}

Last, each test was executed 20 times and the performance was determined based on the average of these multiple runs (to minimize deviations).

\subsection{Single-GPU Performance and Portability Results}

A comprehensive analysis of SYCL and CUDA implementations was conducted on various NVIDIA, AMD, and Intel GPUs to evaluate the performance and portability of SYCL across different GPU architectures. 

\subsubsection{Protein Database Search}

Table~\ref{tab:arch-eff} provides a detailed comparison of the performance and the architectural efficiency of CUDA and SYCL implementations on each platform for protein database search. On NVIDIA GPUs, CUDA and SYCL demonstrated comparable performance and efficiency values. As anticipated, more powerful GPUs achieved higher GCUPS values. The architectural efficiency values ranged from 37\% to 52\%. In particular, while the RTX 3090 GPU exhibited the highest GCUPS value, the RTX 2070 GPU proved to be the most efficient considering architectural usage.

For AMD and Intel GPUs, only SYCL results are presented, as CUDA exclusively supports NVIDIA hardware. This limitation underscores the superior portability of SYCL compared to CUDA. The performance of the SYCL implementation on AMD and Intel GPUs presents a mixed picture. While AMD's RX 6700 XT dGPU achieved efficiency rates comparable to NVIDIA GPUs (51.7\%), the AMD Vega 6 iGPU showed considerably lower performance with only 21.3\% efficiency rate. For Intel GPUs, a notable dichotomy emerged between their graphic architectures. Intel's Gen-based iGPUs (UHD 630 and UHD 770) demonstrated impressive architectural efficiency, reaching up to 75.7\%, significantly surpassing the baseline established by NVIDIA GPUs. However, Intel's Xe-based GPUs (Arc A770 dGPU and Xe-LPG 128EU iGPU) exhibited markedly lower performance, with architectural efficiencies down to 23.4\%, respectively. This suboptimal performance can be attributed to ongoing driver optimization challenges in Intel's Arc generation, a factor that has contributed to Intel's dGPU market share dropping below 1\% in 2024~\cite{norem24}. These performance limitations with Xe-based GPUs will likely impact the overall performance portability metrics for Intel GPUs in this study. Further investigation, including detailed code profiling, is planned to understand these performance discrepancies and their relationship to hardware architecture and driver maturity.

\begin{table}[!t]
\centering
\adjustbox{max width=0.5\textwidth}{
\begin{tblr}{
 cells = {c},
 cell{1}{1} = {c=3}{},
 cell{1}{4} = {c=2}{},
 cell{1}{6} = {c=2}{},
 cell{3}{1} = {r=6}{},
 cell{9}{1} = {r=4}{},
 cell{13}{1} = {r=2}{},
 vlines,
 hline{1,9,13-16} = {-}{},
 hline{4-8,10-12,14-15} = {2-7}{},
}
\textbf{Platform} &              &                                 & \textbf{CUDA} &  & \textbf{SYCL}                                &                                        \\
\rotatebox[origin=c]{90}{\textbf{Vendor}}     & \textbf{GPU} & {\textbf{GCUPS}\\\textbf{peak}} & {\textbf{GCUPS}\\\textbf{ach.}} & {\textbf{Arch.}\\\textbf{eff.}} & {\textbf{GCUPS}\\\textbf{ach.}} & {\textbf{Arch.}\\\textbf{eff.}} \\
\hline
\rotatebox[origin=c]{90}{\textbf{NVIDIA}}            & GTX 980      & 155.5                           & 70.6 & 45.3 & 67.7                                         & 43.5\%                                 \\
                 & GTX 1080     & 277.2                           & 104.5 & 37.7 & 103.8                                        & 37.4\%                                 \\
                 & RTX 2070     & 311.0                           & 162.6 & 52.2 & 163.1                                        & 52.4\%                                 \\
                 & Tesla V100   & 588.8                           & 225.0 & 38.2 & 233.0                                        & 39.6\%                                 \\
                 & RTX 3070     & 423.2                           & 173.2 & 40.9 & 174.4                                        & 41.2\%                                 \\
                 & RTX 3090     & 741.3                           & 280.2 & 37.8 & 288.6                                        & 38.9\%                                 \\
\rotatebox[origin=c]{90}{\textbf{Intel}}             & Arc A770     & 819.2                           & x & NA & 191.4                                        & 23.4\%                                 \\
& UHD 630 & 19.2 & x & NA & 13.1 & 68.4\%                                 \\
& UHD 770      & 35.2 & x & NA & 26.6 & 75.7\%                                 \\
& Xe-LPG 128EU & 192.0 & x & NA & 53.5 & 27.9\%                                    \\
\rotatebox[origin=c]{90}{\textbf{AMD}}               
                 & \begin{tabular}[c]{@{}c@{}}RX 6700\\ XT \end{tabular}   & 550.6  & x & NA & 284.4                                        & 51.7\%                                 \\
                 & RX Vega 6       & 35.2    & x & NA & 7.5                                          & 21.3\%
\end{tblr}
}
\caption{GCUPS and architectural efficiencies of CUDA and SYCL codes on single GPUs  for protein database search.}
\label{tab:arch-eff}
\end{table}

Table~\ref{tab:performance-portability} presents the performance portability of both CUDA and SYCL implementations. The aggregated results are consistent with the individual observations previously noted. For NVIDIA GPUs, the performance portability of CUDA and SYCL is remarkably similar, with values of 42\% and 42.2\%, respectively. In the case of Intel GPUs, SYCL shows distinct performance patterns: excellent architectural efficiency (up to 75.7\%) on Gen-based iGPUs, but significantly lower efficiency (23.4\%) on the Arc-based dGPU. As a consequence, SYCL presents 48.9\% of performance portability on Intel GPUs. The combination of AMD and Intel GPUs yields performance portability of 44.7\%, higher than other vendor combinations. However, including NVIDIA GPUs in the analysis reduces overall performance portability to 43.5\%, primarily due to the lower efficiency rates observed in Intel's Xe-based GPUs and AMD's integrated solutions.

\begin{table}[t]
\begin{center}

\begin{tabular}{|c|c|c|}
\hline
\multicolumn{1}{|c|}{} & \multicolumn{2}{|c|}{$\funnyPbar(\alpha,p,H)$} \\

\textbf{Platform set (\textit{H})} & \textbf{CUDA} & \textbf{SYCL} \\

NVIDIA  &	42\% &	42.2\% \\
AMD & NA & 36.5\% \\
Intel &	NA	& 48.9\% \\ 
\hdashline
NVIDIA $ \cup $ AMD &	NA &	40.8\% \\
NVIDIA $ \cup $ Intel &	NA &	44.9\% \\
Intel $ \cup $ AMD &	NA &	44.7\% \\
\hdashline
NVIDIA $ \cup $ AMD $ \cup $ Intel &	NA &	43.5\% \\
\bottomrule
\end{tabular}
\end{center}
\caption{Performance portability of both CUDA and SYCL codes on single GPUs  for protein database search.}
\label{tab:performance-portability}
\end{table}


\subsubsection{Pairwise Alignment}

Table~\ref{tab:arch-eff-dna} presents the performance and the architectural efficiency of CUDA and SYCL implementations on each platform for pairwise alignment. As in the protein database search case, CUDA and SYCL demonstrated comparable performance and efficiency values on NVIDIA GPUs. Once again, the highest GCUPS value was achieved by the RTX 3090 while the most efficient GPU considering architectural usage was the RTX 2070.

The results on the Intel GPUs are mixed. The Arc A770 achieved almost the same architectural efficiency as the protein benchmark. However, the two Gen-based iGPUs showed lower performance while the XE-LPG 128 EU achieved a higher architectural efficiency than the previous case. This behavior could be related to the datatype used in the pairwise alignment kernel (\textit{long kernel}), as discussed below in Section~\ref{sec:cpu-perf-discuss}.

\begin{table}[!t]
\centering

\adjustbox{max width=0.5\textwidth}{
\begin{tblr}{
 cells = {c},
 cell{1}{1} = {c=3}{},
 cell{1}{4} = {c=2}{},
 cell{1}{6} = {c=2}{},
 cell{3}{1} = {r=5}{},  
 cell{8}{1} = {r=4}{}, 
 vlines,
 hline{1,8,12-15} = {-}{},  
 hline{4-7,9-11,13-14} = {2-7}{}, 
}
\textbf{Platform} &              &                                 & \textbf{CUDA} &  & \textbf{SYCL}                                &                                        \\
\rotatebox[origin=c]{90}{\textbf{Vendor}}     & \textbf{GPU} & {\textbf{GCUPS}\\\textbf{peak}} & {\textbf{GCUPS}\\\textbf{ach.}} & {\textbf{Arch.}\\\textbf{eff.}} & {\textbf{GCUPS}\\\textbf{ach.}} & {\textbf{Arch}\\\textbf{eff.}} \\
\hline
\rotatebox[origin=c]{90}{\textbf{NVIDIA}}    
& GTX 980 & 155.5 & 60.3 & 38.7 & 60.4 & 38.8\% \\
& GTX 1080 & 277.2 & 125.0 & 45.1 & 125.2 & 45.1\%  \\
& RTX 2070 & 311.0 & 197.9 & 63.6 & 218.8 & 70.3\% \\
& Tesla V100 & 588.8 & 257.6 & 43.8 & 278.9 & 47.4\% \\
& RTX 3090 & 741.3 & 487.5 & 65.8 & 496.3 & 67.0\% \\
\rotatebox[origin=c]{90}{\textbf{Intel}}
& Arc A770 & 819.2 & x & NA & 193.6 & 23.6\% \\
& UHD 630 & 19.2 & x & NA & 7.9 & 41.3\% \\
& UHD 770 & 35.2 & x & NA & 14.79 & 42.0\% \\
& Xe-LPG 128EU & 192.0 & x & NA & 83.30 & 43.4\% \\
\end{tblr}
}
\caption{GCUPS and architectural efficiencies of CUDA and SYCL codes on single GPUs for pairwise alignment}
\label{tab:arch-eff-dna}
\end{table}

The performance portability of both CUDA and SYCL implementations for pairwise alignment is presented in Table~\ref{tab:performance-portability-dna}. Once again, the performance portability of CUDA and SYCL is remarkably similar for NVIDIA GPUs. In particular, CUDA and SYCL present values of 51.4\% and 53.7\%, respectively. This parity observed for both protein and DNA alignment indicates that both programming models can deliver consistent performance across the diverse range of NVIDIA GPUs utilized in this study.

In the case of Intel GPUs, while the iGPUs exhibited an acceptable performance, the discrete GPU showed low architectural efficiency values. This fact explains why SYCL presents 37.6\% of performance portability for pairwise alignment. Last, the combination of NVIDIA and Intel GPUs yields performance portability of 45.7\%, a slightly higher value than the one achieved by the same platform set for the protein benchmark.

\begin{table}[t]
\begin{center}  
\begin{tabular}{|c|c|c|}
\hline
\multicolumn{1}{|c|}{} & \multicolumn{2}{|c|}{$\funnyPbar(\alpha,p,H)$} \\
\textbf{Platform set (\textit{H})} & \textbf{CUDA} & \textbf{SYCL} \\
NVIDIA  &	51.4\% &	53.7\% \\
Intel &	NA	& 37.6\% \\ 
\hdashline
NVIDIA $ \cup $ Intel &	NA &	45.7\% \\
\bottomrule
\end{tabular}
\end{center} 
\caption{Performance portability of both CUDA and SYCL codes on single GPUs for pairwise alignment.}
\label{tab:performance-portability-dna}
\end{table}

\subsubsection{Discussion}

Building upon this analysis, SYCL demonstrates comparable performance portability to CUDA throughout this study. Specifically, SYCL achieves nearly equivalent architectural efficiency to CUDA across six NVIDIA GPUs representing five distinct microarchitectures. SYCL's key advantage lies in its cross-vendor compatibility, successfully executing on both AMD and Intel GPUs, with particularly strong performance on Intel iGPUs and AMD dGPUs, despite the lower efficiency observed on Intel Arc and AMD integrated solutions.

This comprehensive evaluation underscores SYCL's broad compatibility and its ability to enhance performance across a diverse range of GPU architectures for the application under consideration. The results highlight SYCL's potential as a versatile solution for heterogeneous computing environments, offering a balance between performance and portability that extends beyond the capabilities of CUDA\cite{Haseeb2021,vasileska24,JinVetter2022_1, JinVetter2022_2,SolisVasquez2023,castano2022evaluation}.

\subsection{Multi-GPU Performance and Portability Results}

Experimental work was conducted on various NVIDIA and Intel multi-GPU setups to further investigate the performance and portability of SYCL when exploiting multiple GPUs~\footnote{The pairwise alignment benchmark was not considered for this analysis because SW\# only support multi-device configuration for protein database search}. Table~\ref{tab:arch-eff-multiple} presents a detailed comparison of the performance and architectural efficiency of CUDA and SYCL implementations across five distinct multi-GPU configurations  for protein database search. 

For NVIDIA multi-GPU configurations, the efficiency rates achieved when utilizing two GPUs in combination are generally slightly lower than those observed with single-GPU configurations. This phenomenon is evident in three of the four tested configurations (except for 2$\times$Tesla V100) and can be attributed to two primary reasons. On the one hand, efficiency usually decreases when the problem size remains fixed while computational resources are increased. On the other hand, the workload distribution strategy employed by \textit{SW\#} is rather simple : it distributes query sequences among GPUs without considering sequence length and individual GPU computing power. Given that these sequences vary in length, this approach can lead to load imbalance between GPUs, potentially reducing overall performance. This situation worsens in heterogeneous multi-GPU environments.

\subsubsection{Discussion}

SYCL once again demonstrates its enhanced functional portability in the case of Intel's multi-GPU configuration. While the performance in this scenario is suboptimal for the previously mentioned reasons, it is noteworthy that SYCL enables the simultaneous utilization of two Intel GPUs of different types: an integrated GPU (iGPU) and a discrete GPU (dGPU). This capability underscores the potential for SYCL to leverage heterogeneous computing resources within a single system.

These findings have significant implications for HPC in heterogeneous environments. The ability of SYCL to maintain performance parity with CUDA in NVIDIA multi-GPU configurations, while also enabling cross-vendor GPU utilization \cite{Haseeb2021, JinVetter2022_1, JinVetter2022_2, SolisVasquez2023, bria20}, positions it as a versatile solution for developers seeking to maximize computational resources across diverse hardware landscapes\cite{ Faqir23, jin23, apa23, torres24, faqir24}. However, the observed efficiency variations in multi-GPU setups highlight the need for more sophisticated workload distribution strategies to fully capitalize on the potential of heterogeneous computing environments, as it was considered in~\cite{schmidt2024cudasw++}.

\begin{table}[!t]
\begin{adjustbox}{width=\columnwidth,totalheight=\textheight,keepaspectratio}
\begin{tblr}{
  cells = {c},
  cell{1}{1} = {c=3}{},
  cell{1}{4} = {c=2}{},
  cell{1}{6} = {c=2}{},
  cell{3}{1} = {r=4}{},
  cell{7}{1} = {r=1}{},
  vlines,
  hline{1,7,8} = {-}{},
  hline{4-6} = {2-7}{},
}
\textbf{Platform} &              &                                 & \textbf{CUDA}                                &                                            & \textbf{SYCL}                                &                                        \\
\rotatebox[origin=c]{90}{\textbf{Vendor}}     & \textbf{GPUs} & {\textbf{GCUPS}\\\textbf{peak}} & {\textbf{GCUPS}\\\textbf{\textbf{ach.}}} & {\textbf{Arch.}\\\textbf{\textbf{eff.}}} & {\textbf{GCUPS}\\\textbf{\textbf{ach.}}} & {\textbf{Arch.}\\\textbf{\textbf{eff.}}} \\
\hline

\rotatebox[origin=c]{90}{\textbf{NVIDIA}} & \begin{tabular}[c]{@{}c@{}}$2\times$ \\ GTX 1080\end{tabular} & 554.6 & 189.8 & 34.2\% & 187.8 & 33.9\% \\
                  & \begin{tabular}[c]{@{}c@{}}$2\times$ \\ Tesla V100 \end{tabular} & 1177.6 & 306.5 & 36.2\% & 308.9 & 36.5\% \\
                  & \begin{tabular}[c]{@{}c@{}}$2\times$ \\ RTX 3070 \end{tabular}& 846.4 & 318.1 & 27.0\% & 336.5 & 28.6\% \\
                  & \begin{tabular}[c]{@{}c@{}} Tesla V100 \\ +\\ RTX 3090\end{tabular} & 1330.1 & 450.5 & 33.8\% & 460.7 & 34.6\% \\

\rotatebox[origin=c]{90}{\textbf{Intel}} & \begin{tabular}[c]{@{}c@{}} Arc A770 \\ +  \\ UHD 770 \end{tabular} & 854.4 & $\times$ & NA & 126.8 & 14.8\% \\
\end{tblr}
\end{adjustbox}

\caption{GCUPS and architectural efficiencies of both CUDA and SYCL codes on multiple GPUs  for protein database search.}
\label{tab:arch-eff-multiple}
\end{table}

\subsection{CPU Performance and Portability Results}
\label{perf-individual-cpu}

A comprehensive analysis was conducted on various Intel and AMD processors to evaluate the portability and performance of SYCL on different CPU architectures,

\subsubsection{Protein Database Search}

\begin{figure}[!t]
 \centering
      \includegraphics[width=1\columnwidth]{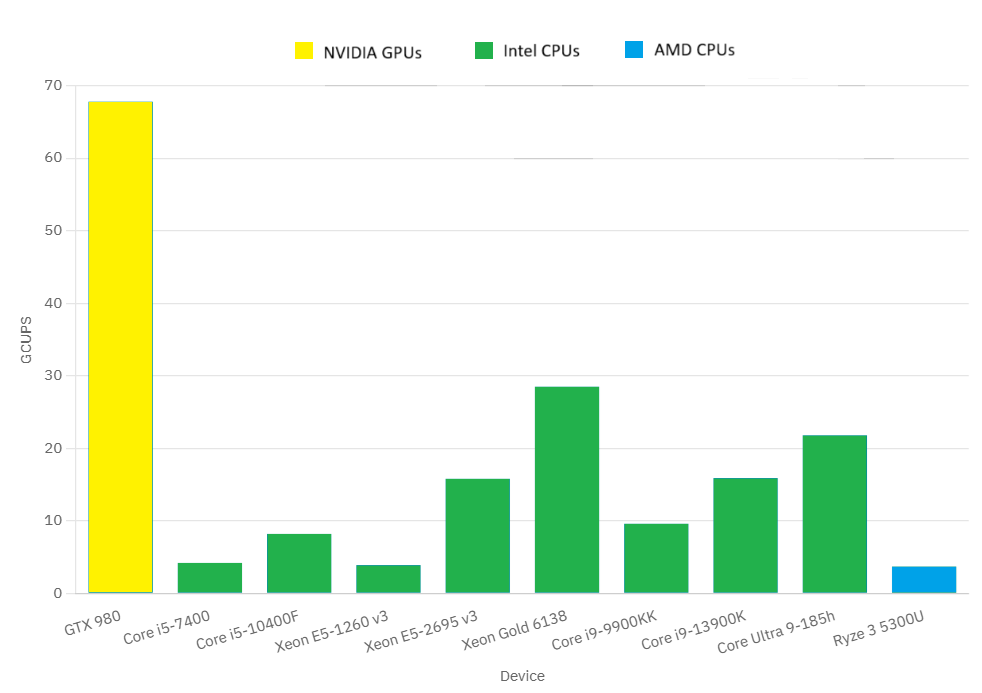}
     \caption{Performance comparison between Intel CPUs and AMD CPU  for protein database search, using an NVIDIA GPU as reference.}
     \label{fig:individual-cpus}     
 \end{figure}

Fig.~\ref{fig:individual-cpus} presents a performance comparison between Intel and AMD CPUs, comparing them with an NVIDIA GPU as a reference. It is important to remark that the main objective of this figure is to showcase the adaptability of SYCL code across diverse CPU architectures. The SYCL code demonstrates remarkable adaptability by successfully executing on a diverse range of CPUs from different manufacturers, including the Intel Core i5-7400 and AMD Ryzen 3 5300U models. Notably, the code also ran effectively on the Core i9-13900K and Core Ultra 9-185H CPUs, which employ a hybrid architecture.

Even though the original, CUDA code was designed for NVIDIA GPUs, the migrated SYCL code could be adapted and executed on CPUs from various manufacturers with minimal modifications. Two key points deserve emphasis: first, running these tests only required changing the backend during compilation; second, since the ported DPC++ version is pure SYCL code, executing this version on other architectures will only necessitate a compatible compiler.

While functional portability to CPUs has been established, examining the corresponding performance portability is important. Table~\ref{tab:arch-eff-cpu} presents the GCUPS performance and architectural efficiency of the SYCL code running on CPUs for protein database search. The Core i5-10400F  emerges as the top performer among all CPUs, achieving a maximum efficiency of 51.2\%. Conversely, the Core i9-13900K exhibits the lowest architectural efficiency despite boasting the highest theoretical peak. It's noteworthy that while CPUs generally deliver lower GCUPS compared to GPUs, architectural efficiency surpasses 38\% in 7 out of 9 instances.

To provide a more comprehensive analysis, Table~\ref{tab:performance-portability-cpu} presents the performance portability on CPUs, categorizing results by manufacturers and segments~\footnote{Single-member groups are discarded}. Among Intel CPUs, the desktop segment demonstrates a slight advantage over the server segment. A similar trend is observed when comparing different manufacturers, with Intel CPUs showing marginally better performance than the selected AMD CPU. It is important to note that this discrepancy could be attributed to the fact that the oneAPI ecosystem is developed by Intel, which may result in more efficient code generation for their processors. However, caution should be taken in drawing broad conclusions, given the limited number of CPUs within each platform set.

The $\funnyPbar$ metric offers a means to assess the adaptability of an application when transitioning between architectures. It's crucial to recognize that \textit{SW\#} is a CUDA code specifically optimized for NVIDIA GPUs, while its SYCL counterpart achieved a $\funnyPbar$ value of 42.2\% during testing on 6 distinct GPUs with 5 different microarchitectures. According to Table~\ref{tab:performance-portability-cpu}, the performance portability across all CPU segments and manufacturers stands at 40.3\%, exhibiting a negligible difference of less than 2\%. This observation underscores the effectiveness of SYCL code across various platforms and highlights its advantage in terms of portability, a critical factor in environments characterized by diverse hardware availability and flexible hardware selection.

\begin{table}[!t]
\centering
\begin{tblr}{
cells = {c},
  cell{1}{1} = {c=3}{},
  cell{1}{4} = {c=2}{},
  cell{3}{1} = {r=8}{},
  cell{11}{1} = {r=1}{},
  vlines,
  hline{1,3,11,12} = {-}{},
  hline{4-11} = {2-5}{},
}
\textbf{Platform} &              &                                 & \textbf{SYCL}                                &                                        \\
\rotatebox[origin=c]{90}{\textbf{Vendor}}     & \textbf{CPU} & {\textbf{GCUPS}\\\textbf{peak}} & {\textbf{GCUPS}\\\textbf{\textbf{ach.}}} & {\textbf{Arch.}\\\textbf{\textbf{eff.}}} \\
\rotatebox[origin=c]{90}{\textbf{Intel}} 
 & {Core \\i5-7400} & 8.8 & 4.2 & 47.6\% \\
 & {Core \\i5-10400F} & 16.0 & 8.2 & 51.2\% \\
 & {Xeon \\E5-1620 v3} & 9.3 & 3.9 & 42.1\% \\
 & {Xeon \\E5-2695 v3} & 35.5 & 15.8 & 44.6\% \\
 & {Xeon \\Gold 6138} & 101.3 & 28.5 & 28.2\% \\
 & {Core \\i9-9900K} & 25.1 & 9.6 & 38.4\% \\
 & {Core \\i9-13900K} & 75.2 & 15.9 & 21.1\% \\
 & {Core \\U9-185H} & 44 & 21.8 & 49.5\% \\
\rotatebox[origin=c]{90}{\textbf{AMD}} 
 & {Ryzen 3 \\5300U}& 9.6 & 3.7 & 38.8\%
\end{tblr}
\caption{GCUPS and architectural efficiencies of SYCL code on individual CPUs  for protein database search.}
\label{tab:arch-eff-cpu}
\end{table}

\begin{table}[htbp]
\begin{center}
\begin{tabular}{|c|c|}
\hline
\multicolumn{1}{|c|}{} & \multicolumn{1}{|c|}{$\funnyPbar(\alpha,p,H)$} \\

\textbf{Platform set (\textit{H})} & \textbf{SYCL} \\
Intel CPUs (desktop) & 39.6\% \\
Intel CPUs (server)  & 38.3\% \\
Intel CPUs & 40.3\% \\
\hdashline
Intel $ \cup $ AMD  & 40.3\% \\
\bottomrule
\end{tabular}
\end{center}

\caption{Performance portability of SYCL code on individual CPUs  for protein database search.}
\label{tab:performance-portability-cpu}
\end{table}

\subsubsection{Pairwise Alignment}

Table~\ref{tab:arch-eff-cpu-dna} presents the GCUPS performance and architectural efficiency of the SYCL code running on CPUs for pairwise alignment. Unlike the protein database search case, SYCL achieves significantly lower architectural efficiency values for CPUs than GPUs. Through code review and profiling, we attribute this difference to the fact that the kernel for long sequences uses scalar data instead of vector data. For protein benchmark, \textit{SW\#} uses the \textit{short kernel} for almost all alignments, which leads to explicit instruction vectorization. This code change effectively impacts performance on the CPUs, as the compiler is not able to vectorize operations, achieving lower GCUPS. Moreover, the computation of performance portability reflects the previous behavior. On the one hand, SYCL presents practically the same values of $\funnyPbar$ for the protein benchmark (43.5\% on GPUs vs 40.2\% on CPUs). On the other hand,  when pairwise alignment is considered, SYCL achieves $\funnyPbar$ values of 45.7\% and 11.5\% on GPUs and CPUs, respectively. Although the platform sets are similar but not identical, the difference is large enough to notice the performance losses.

\begin{table}[!t]
\centering  
\begin{tblr}{
cells = {c},
  cell{1}{1} = {c=3}{},
  cell{1}{4} = {c=2}{},
  cell{3}{1} = {r=4}{},  
  cell{7}{1} = {r=1}{},  
  vlines,
  hline{1,3,4,5,6,7,8} = {-}{},  
}
\textbf{Platform} &              &                                 & \textbf{SYCL}                                &                                        \\
\rotatebox[origin=c]{90}{\textbf{Vendor}}     & \textbf{CPU} & {\textbf{GCUPS}\\\textbf{peak}} & {\textbf{GCUPS}\\\textbf{\textbf{ach.}}} & {\textbf{Arch.}\\\textbf{\textbf{eff.}}} \\
\rotatebox[origin=c]{90}{\textbf{Intel}} 
 & {Core \\i5-10400F} & 16.0 & 2.0 & 12.2\% \\
 & {Xeon \\E5-1620 v3} & 0.9 & 3.9 & 9.7\% \\
 & {Xeon \\Gold 6138} & 101.3 & 17.38 & 17.2\% \\
 & {Core \\i9-13900K} & 75.2 & 3.8 & 5.1\% \\
\end{tblr}
\caption{GCUPS and architectural efficiencies of SYCL code on individual CPUs for pairwise alignment}
\label{tab:arch-eff-cpu-dna}
\end{table}

\subsubsection{Discussion}
\label{sec:cpu-perf-discuss}

These findings have significant implications for the development of high-performance, portable applications in fields such as bioinformatics. 
SYCL demonstrated some ability to maintain consistent performance across diverse CPU architectures~\cite{Haseeb2021,Faqir23,reguly23,range23}. However, kernel programming plays an important role in achieving that goal and programmers should consider this issue if they plan to execute their code on both CPUs and GPUs. In addition, as time goes by, compilers will improve their vectorization capabilities to cope with the current limitations. In that sense, SYCL could serve as a powerful tool for researchers and developers seeking to optimize computational performance while maintaining code portability~\cite{castano2022evaluation,faqir24,range23} in heterogeneous computing environments.

\subsection{CPU-GPU Performance and Portability Results}
\label{perf-cpu-gpu}
To investigate the performance and portability of SYCL in hybrid CPU-GPU configurations, experiments were conducted on various combinations of NVIDIA, Intel, and AMD devices~\footnote{The pairwise alignment benchmark was not considered for this analysis because \textit{SW\# only supports multi-device configuration for protein database search}}. Table~\ref{tab:cpu-gpu-efficiency} provides a comprehensive overview of GCUPS and architectural efficiencies for codes executed on various CPU-GPU combinations using SYCL  for protein database search~\footnote{Single-member groups are discarded}. As previously noted, CUDA's unsuitability for CPUs further accentuates the significance of SYCL in hybrid computing scenarios. Analysis of the table reveals that combinations involving the GTX 980 (Maxwell) and RTX 3070 (Ampere) GPUs demonstrate the highest performance, highlighting the code's scalability across different NVIDIA architectures. Nevertheless, even configurations with lower GCUPS, such as Vega 6 and Ryzen 3 5300U, showcase functional portability across a broader spectrum of hardware, including those with more constrained resources.

The architectural efficiency of SYCL exhibits variability across different CPU-GPU combinations, illustrating the code's adaptability to the unique characteristics of each architecture. Configurations featuring NVIDIA GPUs and Intel CPUs demonstrate higher architectural efficiency at the upper end of the table, suggesting a more seamless integration between these technologies. In contrast, pairings of Intel GPUs and CPUs, as well as AMD CPUs and GPUs, display lower architectural efficiencies, highlighting potential areas for enhancing the synergy between these architectures.

Table~\ref{tab:performance-portability-cpu-gpu} presents an analysis of performance portability across various hybrid combinations  for protein database search. Optimal performance with SYCL is observed in configurations that combine an Intel CPU and an NVIDIA GPU, reaching 27.3\%, suggesting potentially more efficient optimization or enhanced compatibility between these specific devices to execute the given codes. Conversely, the least favorable performance is noted with an Intel CPU and an AMD dGPU combination, achieving only 6\%.

Configurations incorporating AMD and Intel iGPUs demonstrate moderate performance levels, ranging from 14.03\% to 17.8\%, respectively. These findings underscore the significant impact of the selection of the CPU and GPU models on the performance of the SYCL code, highlighting the importance of application optimization in integrated CPU+GPU systems. To enhance the validity of these findings, future studies would benefit from an expanded collection of GPUs, particularly from Intel and AMD, as well as an increased number of CPUs, especially from AMD.

\subsubsection{Discussion}

In this context, incorporating additional devices that present different computing power generally results in performance degradation, even below individual rates. Similar to multi-GPU experiments, outcomes are predominantly influenced by the \textit{SW\#} workload distribution strategy, which fails to account for individual device capabilities. Consequently, the \textit{slow device} may be burdened with aligning longer sequences compared to the \textit{fast one, which will remain idle the rest of the time. This situation is independent of which device is fast or slow. For example, when using the Core i9-13900K+UHD770 configuration, the CPU is the fast device and the GPU is the slow one. However, the opposite case occurs when using the Ryzen 3 5300U+RX Vega 6 system. In any case, the load imbalance detrimentally impacts} overall performance. Furthermore, although not equivalent to utilizing two GPUs simultaneously, CPU-GPU combinations are also affected when the problem size remains constant while computational resources increase. This scenario presents potential avenues for future optimizations, including the development of more sophisticated workload distribution algorithms.

\begin{table}[!t]
\centering
\begin{tblr}{
  cells = {c},
  cell{1}{1} = {c=3}{},
  cell{1}{4} = {c=2}{},
  cell{3}{1} = {r=6}{},
  cell{9}{1} = {r=4}{},
  cell{13}{1} = {r=1}{},
  cell{14}{1} = {r=1}{},
  vlines,
  hline{1,3,9,12,13,14,15} = {-}{},
  hline{4-9,10-14} = {2-5}{}
}
\textbf{Platform} &              &                                 & \textbf{SYCL}                                &                                        \\
\rotatebox[origin=c]{90}{\textbf{Conf.}}     & {\textbf{CPU}\\\textbf{+}\\\textbf{GPU}}  & {\textbf{GCUPS}\\\textbf{peak}} & {\textbf{GCUPS}\\\textbf{\textbf{ach.}}} & {\textbf{Arch.}\\\textbf{\textbf{eff.}}} \\
\hline
\rotatebox[origin=c]{90}{\textbf{\begin{tabular}{@{}c@{}}Intel CPU\\+\\NVIDIA GPU\end{tabular}}} 
                 & \begin{tabular}{@{}c@{}} Xeon E5-2695 v3\\+\\GTX980\end{tabular} & 191 & 73 & 38.1\% \\
                 & \begin{tabular}{@{}c@{}}Xeon E5-2695 v3\\+\\GTX1080\end{tabular} & 287 & 84 & 29.5\% \\
                 & \begin{tabular}{@{}c@{}}Core i5-7400\\+\\RTX2070\end{tabular} & 320 & 49 & 15.3\% \\
                 & \begin{tabular}{@{}c@{}}Core i5-10400F\\+\\RTX3070\end{tabular} & 439 & 149 & 33.9\% \\
                 & \begin{tabular}{@{}c@{}}Xeon Gold 6138\\+\\RTX3090\end{tabular} & 843 & 210 & 24.9\% \\
                 & \begin{tabular}{@{}c@{}}Xeon Gold 6138\\+\\V100\end{tabular} & 690 & 156 & 22.6\% \\
\rotatebox[origin=c]{90}{\textbf{\begin{tabular}{@{}c@{}}Intel CPU\\+\\Intel GPU\end{tabular}}} 
                 & \begin{tabular}{@{}c@{}}Core i9-13900K\\+\\Arc A770\end{tabular} & 894 & 124 & 13.9\% \\
                 & \begin{tabular}{@{}c@{}}Core i9-13900K\\+\\UHD770\end{tabular} & 110 & 13 & 12\% \\
                 & \begin{tabular}{@{}c@{}}Core i9-13900K\\+\\Arc A770\\+\\UHD770\end{tabular} & 930 & 102 & 11\% \\
                 & \begin{tabular}{@{}c@{}}Core U9-185H\\+\\Xe-LPG 128EU\end{tabular} & 236 & 25.8 & 10.9\% \\
\rotatebox[origin=c]{90}{\textbf{\begin{tabular}{@{}c@{}}Intel CPU\\+\\AMD GPU\end{tabular}}} 
                 & \begin{tabular}{@{}c@{}}Xeon E5-1620 v3\\+\\RX6700\end{tabular} & 560 & 33 & 6\%  \\
\rotatebox[origin=c]{90}{\textbf{\begin{tabular}{@{}c@{}}AMD CPU\\+\\AMD GPU\end{tabular}}} 
                 & \begin{tabular}{@{}c@{}}Ryzen 3 5300U\\+\\RX Vega 6\end{tabular} & 45.0 & 6 & 14.0\%  \\
\end{tblr}%
\caption{GCUPS and architectural efficiencies of CPU+GPU codes on multiple combinations  for protein database search.}
\label{tab:cpu-gpu-efficiency}
\end{table}

\begin{table}[htbp]
\begin{center}
\begin{tabular}{|c|c|}
\hline
\multicolumn{1}{|c|}{} & \multicolumn{1}{|c|}{$\funnyPbar(\alpha,p,H)$} \\

\textbf{Platform set (\textit{H})} & \textbf{SYCL} \\

Intel CPU + NVIDIA GPU  & 27.4\% \\
Intel CPU + Intel GPU  & 12\% \\
\hdashline
Intel CPU + GPU(s)  & 19.8\% \\
\hdashline
CPU + GPU(s)  & 19.3\% \\
\bottomrule
\end{tabular}
\end{center}

\caption{Performance portability of SYCL code on CPU+GPU  for protein database search.}

\label{tab:performance-portability-cpu-gpu}
\end{table}

\section{Related Works}
\label{sec:relworks}
Several studies have compared the performance of SYCL and CUDA implementations across various domains and architectures. These studies can be categorized based on the focus of their comparisons and the architectures used:
\begin{itemize}
    \item  Performance comparison on NVIDIA GPUs: In~\cite{Haseeb2021, JinVetter2022_1, JinVetter2022_2, SolisVasquez2023, bria20, Faqir23, jin23, apa23, torres24, faqir24, vasileska24}, the authors compared SYCL and CUDA implementations on NVIDIA GPUs, finding that SYCL performance is generally comparable to CUDA, with some cases showing a performance gap attributed to differences in compiler optimizations and memory management.
\item Performance comparison on Intel and AMD GPUs:
Studies in~\cite{Haseeb2021, SolisVasquez2023, Faqir23, reguly23, range23, vasileska24} evaluated SYCL performance on Intel and AMD GPUs, demonstrating that SYCL can achieve equivalent or superior performance compared to native programming approaches on these architectures.

\item Performance comparison on CPUs and FPGAs:
The works in~\cite{reguly23, range23, franquinet23, decastro24} investigated SYCL performance on various CPU architectures, finding that SYCL is generally less efficient on CPUs compared to native approaches, highlighting the need for additional optimizations.

\item Performance portability and migration experiences:
Several papers~\cite{JinVetter2022_1, range23, franquinet23, nguyen23, carratal23, jin23_2, weckert23, mueller24} focused on the process of migrating applications from CUDA to SYCL and the performance portability achieved across different architectures. These studies emphasize the importance of device-specific optimizations and parallelization approaches to maximize SYCL performance.

\item Domain-specific applications:
Some studies focused on specific application domains, such as bioinformatics~\cite{Haseeb2021, JinVetter2022_1, JinVetter2022_2, jin23}, cosmology~\cite{range23}, iterative methods~\cite{nguyen23}, FTLE calculations~\cite{carratal23}, and molecular dynamics simulations~\cite{apa23}. These articles provide insights into the performance and portability of SYCL in these particular fields.
\end{itemize}

The studies mentioned above have significantly contributed to understanding the performance and portability of SYCL across various architectures and application domains. However, most of these studies focused on comparing SYCL with native programming models, such as CUDA, on a limited set of platforms, primarily GPUs. Building upon our previous works~\cite{Costanzo2022IWBBIO,costanzo23,supercompcost23}, which evaluated SYCL's performance for the SW\# application on GPUs and multi-GPU configurations, this study presents the most comprehensive assessment of SYCL's functional and performance portability to date, encompassing an unprecedented range of computing architectures. Our analysis extends across 12 different GPUs (6 NVIDIA, 4 Intel, and 2 AMD), 9 distinct CPU models (8 Intel and 1 AMD) from various market segments (desktop, server, and mobile), and multiple hybrid CPU-GPU configurations. This extensive hardware coverage, which significantly surpasses previous studies in the field, allowed us to thoroughly investigate SYCL's capabilities across diverse architectural paradigms, from traditional single-device setups to complex heterogeneous systems combining CPUs and GPUs from different vendors.


\section{Conclusions and Future Work}
\label{sec:conc}

This study addresses the challenges of functional and performance portability in heterogeneous computing environments by evaluating SYCL and CUDA for Smith-Waterman biological sequence alignment across diverse GPU and CPU architectures, including multi-GPU and hybrid configurations. The experimental results demonstrate that CUDA and SYCL deliver comparable performance on NVIDIA GPUs in both single and multi-GPU scenarios, while SYCL exhibits superior architectural efficiency on AMD and Intel GPUs in the majority of cases tested. In multi-GPU configurations, both programming models showed similar scaling behavior. While the performance improved when using additional GPUs, the current workload distribution strategy somewhat limited the associated benefits. These findings underscore SYCL's broad compatibility and its potential to enhance performance across a diverse range of GPU configurations for this specific application, from single accelerators to more complex multi-GPU setups.

SYCL also demonstrated remarkable versatility and effectiveness across CPUs from various manufacturers, including the latest hybrid architectures like Intel's Meteor Lake, which combines up to three different types of cores within a single processor. This successful adaptation to both traditional and modern heterogeneous CPU designs is particularly noteworthy, as it showcases SYCL's ability to handle increasingly complex processor architectures effectively. Even so, achieving performance portability to CPUs can remain challenging in some cases, particularly when instruction vectorization cannot be guaranteed. However, the success stories are promising, as the protein benchmark case. The performance consistency across all CPU categories and manufacturers was marginally lower than that achieved on NVIDIA GPUs, highlighting SYCL's potential as a unified programming model for heterogeneous computing environments, regardless of the underlying architectural complexity.

Furthermore, SYCL showed exceptional functional portability in various CPU-GPU hybrid configurations. However, performance in these heterogeneous scenarios was constrained by limitations inherent in the original code base rather than SYCL-specific issues. This highlights the importance of developing more sophisticated work distribution mechanisms that consider the distinct characteristics and capabilities of each computing device in heterogeneous environments. Such optimizations would be crucial for maximizing performance in CPU-GPU configurations, regardless of the programming model used.

Future work will focus on:
\begin{itemize}
    \item Optimizing the SYCL code to achieve maximum performance. Specifically, the original SW\# suite does not incorporate certain known optimizations for SW alignment~\cite{swipe11}, such as instruction reordering to reduce the instruction count and the use of lower-precision integers to increase parallelism~\footnote{It is important to note that at the time of SW\#'s development, most CUDA-enabled GPUs did not support efficient arithmetic operations on 8-bit vector data types.}. Additionally, our objective is to improve the workload distribution strategy when utilizing multiple devices. These improvements are expected to lead to higher efficiency rates.
    
    \item Executing the SYCL code on other FPGA architectures and considering alternative programming models, such as Kokkos and RAJA, to strengthen the current performance portability study. This expansion of scope will provide a more comprehensive understanding of SYCL's capabilities across diverse hardware and software ecosystems, further informing its potential as a unified programming model for heterogeneous computing environments.

\end{itemize}

\section*{CRediT authorship contribution statement}
\textbf{Manuel Costanzo}: Conceptualization, Software, Validation, Writing - Original Draft, Investigation, Visualization, Data Curation. 
\textbf{Enzo Rucci}: Conceptualization, Methodology, Investigation,
Supervision, Writing - Original Draft, Project administration. 
\textbf{Carlos García-Sanchez}: Conceptualization, Investigation, Writing - Review \& Editing, Resources.
\textbf{Marcelo Naiouf}: Writing - Review \& Editing, Funding acquisition.
\textbf{Manuel Prieto-Matías}: Writing - Review \& Editing, Funding acquisition.

\section*{Declaration of competing interest}
The authors declare that they have no known competing financial interests or personal relationships that could have appeared to
influence the work reported in this paper.

\section*{Declaration of generative AI and AI-assisted technologies in the writing process}
During the preparation of this work the authors used ChatGPT in order to improve language and readbility. After using this tool/service, the authors reviewed and edited the content as needed and take full responsibility for the content of the publication.

\section*{Data availability}

The SW\# CUDA software used in this study is available at \url{https://github.com/mkorpar/swsharp}, while the migrated SW\# software is available at \url{https://github.com/ManuelCostanzo/swsharp_sycl}. The protein data utilized for this research are sourced from the UniProtKB/Swiss-Prot (Swiss-Prot) database (release 2022\_07) (available at \url{https://www.uniprot.org/downloads}), and the Environmental Non-Redundant (Env. NR) database (release 2021\_04) (available at \url{https://ftp.ncbi.nlm.nih.gov/blast/db/}).

\section*{Funding}

Grant PID2021-126576NB-I00 funded by MCIN/AEI/ \\ 10.13039/501100011033 and, as appropriate, by “ERDF A way of making Europe”, by the “European Union” or by the “European Union Next Generation EU/PRTR”.

\bibliographystyle{elsarticle-num} 
\bibliography{elsarticle-template-harv.bib}

\end{document}